\documentclass[11pt,a4paper]{article}
\usepackage{jcappub}
\usepackage{bm}
\newcommand{\bea}{\begin{eqnarray}}
\newcommand{\ena}{\end{eqnarray}}
\newcommand{\Mp}{M_{\rm Pl}}

\def\dd{\mathrm{d}}
\def\mcP{\mathcal{P}}
\def\mcT{\mathcal{T}}
\def\Mpl{M_{\rm Pl}}

\usepackage{amsmath}

\begin{document}

\begin{flushright} {\footnotesize YITP-19-86, IPMU19-0129}  \end{flushright}

\title{Primordial Tensor Non-Gaussianity from Massive Gravity}
\author[a,b]{Tomohiro Fujita,}
\author[c]{Shuntaro Mizuno}
\author[d,e]{and Shinji Mukohyama}
\affiliation[a]{Department of Physics, Kyoto University, Kyoto, 606-8502, Japan}
\affiliation[b]{D\'epartment de Physique Th\'eorique and Center for Astroparticle Physics, \\
Universit\'e de Gen\`eve, Quai E.
Ansermet 24, CH-1211 Gen\`eve 4, Switzerland}
\affiliation[c]{
Department of Liberal Arts and Engineering Sciences,\\
National Institute of Technology, Hachinohe college, Aomori 039-1192, Japan}
\affiliation[d]{
Center for Gravitational Physics, Yukawa Institute for Theoretical Physics,\\ Kyoto University, Kyoto 606-8502, Japan}
\affiliation[e]{Kavli Institute for the Physics and Mathematics of the Universe (WPI), \\
The University of Tokyo Institutes for Advanced Study,\\
The University of Tokyo, Kashiwa, Chiba 277-8583, Japan}

\abstract{
We calculate tensor bispectrum in a theory of massive tensor gravitons which predicts blue-tilted and largely amplified primordial gravitational waves.
We find that a new 3-point interaction can produce a larger tensor bispectrum than the conventional one from 
general relativity.
The bispectrum peaks at the squeezed configuration and its slope towards the squeezed limit is determined by the graviton mass.
This squeezed tensor bispectrum may be observed as the quadrupolar modulation of the tensor power spectrum by interferometers.
}
\maketitle
\section{Introduction}
\label{sec:Introduction}

Cosmic inflation is widely believed to be the most plausible explanation 
for the origin of primordial perturbations in our Universe,
while it is still nontrivial to embed inflation into more fundamental theories 
(see \cite{Baumann:2014nda}, for a review).
Inflation predicts the existence of primordial gravitational waves (GWs),
whose power spectrum amplitude depends on the value of the Hubble expansion rate
during inflation (see \cite{Maggiore:1999vm}, for a review).
In order to detect or put constraints on the primordial GWs, there are future/ongoing
experiments of Cosmic Microwave Background (CMB) B-mode polarization like
Lite BIRD \cite{Matsumura:2013aja} as well as interferometers such as
LISA~\cite{AmaroSeoane:2012je},  
Advanced-LIGO (A-LIGO)~\cite{TheLIGOScientific:2016dpb}
 and 
DECIGO~\cite{Seto:2001qf,Kawamura:2011zz}.
In most models of inflation, the power spectrum of primordial GWs is
almost scale invariant, with a slightly red tilt, where the most promising way to detect 
the primordial GWs is through CMB B-mode polarization.
Nevertheless, it should be stressed that even if inflation occurred, there are scenarios 
where the amplitude of primordial GWs can be amplified at scales much smaller than CMB's
(see for reviews, \cite{Guzzetti:2016mkm,Bartolo:2016ami}).
These scenarios are observationally interesting, since they open up the possibility that
the primordial GWs can be detected by interferometer experiments, even if their signal 
is not observed at the CMB scale.%
\footnote{However, recently a new method was proposed to constrain the amplitude of the primordial GWs
from the CMB data at Mpc scale, see~\cite{Namikawa:2019tax}.}

Among the various scenarios showing interesting features for the primordial GWs at small scales,
massive gravity (see \cite{deRham:2014zqa}, for a review) attracts conspicuous attention
and has been applied to the study on the primordial GWs~\cite{Dubovsky:2009xk, Gumrukcuoglu:2012wt,Fasiello:2015csa, Kuroyanagi:2017kfx}.
In this context, recently we have proposed a new scenario predicting blue-tilted and largely
amplified primordial GWs~\cite{FKMM}.
This prediction is based on the two assumptions,
where the first one is that the mass of tensor graviton is comparable
to the Hubble expansion rate during the inflation and 
the second one is that the mass diminishes to a small value at a certain time during radiation dominated era. 
Owing to the first assumption, the power spectrum of primordial GWs is blue at the end of inflation.
From the second one,  after inflation  until the mass diminishes to a small value,
gravitons are diluted as non-relativistic matter and hence their amplitude can be substantially amplified
compared to the conventional massless gravitons which decay as radiation.
In conventional massive gravity theories including the one proposed by de Rham-Gabadadze-Tolley
(dRGT)~\cite{deRham:2010ik,deRham:2010kj}, it is well known that the tensor graviton mass
that is positive and comparable to the Hubble expansion rate is prohibited around de Sitter background
by the so-called Higuchi bound~\cite{Higuchi:1986py}. Contrary to this, however, there are viable theories
in which the graviton mass in such region does not introduce instabilities,
like the minimal theory of massive gravity (MTMG)~\cite{DeFelice:2015hla,DeFelice:2015moy}.
In MTMG, there are only two physical degrees of freedom propagating as in general relativity
in the gravity sector and  the would be ghost mode is removed from the construction,
where the assumption of the Lorentz-invariance is relaxed
(see~\cite{ArkaniHamed:2003uy,Rubakov:2004eb,Dubovsky:2004sg,Blas:2009my,Comelli:2013txa,Langlois:2014jba}
for works considering interesting phenomenology based on 
Lorentz-violating massive gravity).

The scenario of~\cite{FKMM} 
has similarity with the one of~\cite{Ricciardone:2017kre}
based on 
a generalization of solid inflation~\cite{Gruzinov:2004ty, Endlich:2012pz} dubbed supersolid inflation~\cite{Nicolis:2013lma} (see~\cite{Cannone:2014uqa,Bartolo:2015qvr,Ricciardone:2016lym}
for related works).  Supersolid inflation is a scenario that simultaneously breaks time reparameterization
and spatial diffeomorphisms during inflation based on the Effective Field Theory (EFT) of inflation~\cite{Cheung:2007st}.
Such a symmetry breaking pattern is accompanied by the appearance of the tensor graviton mass
that can be comparable to the Hubble expansion rate during inflation without introducing ghost instability.
Therefore, the fact that the graviton mass comparable to the Hubble expansion rate during inflation
makes the primordial GWs  blue-tilted holds in both scenarios.
The mechanism to enhance the primordial GWs so that it is detectable by interferometers, however,
is different. Since supersolid inflation is based on EFT of inflation described by four scalar fields
having time- and space-dependent $vevs$ which break reparameterization symmetries
of the background, at the end of inflation, these fields are regarded to arrange themselves so that
the space-reparameterization symmetry is recovered. Therefore, in the scenario of~\cite{Ricciardone:2017kre},
the mass of tensor graviton becomes zero at the end of inflation, where the enhancement mechanism 
in the scenario of~\cite{FKMM} is not applicable.%
\footnote{In the scenario of~\cite{Ricciardone:2017kre},  it is possible to obtain
the primordial GWs detectable by interferometers by considering small speed of sound
for the tensor perturbations.}

Given the situation that there are several scenarios that predict primordial GWs detectable
at interferometer scales, it is important to think about how to distinguish them.
In this respect, the statistical property of primordial GWs is a very helpful tool,
as in the case of primordial curvature perturbations.
For example, the primordial GWs from vacuum fluctuations of the metric are 
almost Gaussian~\cite{Maldacena:2002vr,Maldacena:2011nz}.
Furthermore, stochastic gravitational wave backgrounds due to a combination of a large number of 
uncorrelated astrophysical sources is Gaussian to a high degree, due to the central limit theorem.
Therefore, in this paper, we calculate the bispectrum of primordial GWs,
which is the lowest order statistics providing information on non-Gaussianity
of tensor fluctuations, depending on not only the amplitude, but also the shape of the triangle
composed of the three momentum vectors in the scenario of~\cite{FKMM}  
(for other works discussing tensor non-Gaussianity, see 
~\cite{Soda:2011am,Shiraishi:2011st,Gao:2011vs,Huang:2013epa,Zhu:2013fja,Cook:2013xea,Shiraishi:2013kxa,Akita:2015mho,Agrawal:2017awz,Agrawal:2018mrg,Dimastrogiovanni:2018gkl}).

 It seems that the simplest way to detect the primordial tensor bispectrum is the direct measurement
 of the bispectrum at interferometer scales \cite{Bartolo:2018qqn}. However, it was shown that
 the bispectrum cannot be probed directly ~\cite{Bartolo:2018evs,Bartolo:2018rku}.
 This is caused by the fact that
 the tensor bispectrum at small scales is suppressed due to Shapiro time-delay effects associated with
 the propagation of tensor modes at sub-horizon scales in the presence of matter.
 On the other hand,  for supersolid inflation, another method to prove the tensor bispectrum
 based on the quadrupolar modulation of the tensor power spectrum induced by the tensor bispectrum
 was proposed~\cite{Ricciardone:2017kre}.%
 \footnote{The idea that the squeezed tensor bispectrum induces the quadrupolar modulation of the power spectrum 
 of curvature perturbation was proposed earlier and has been investigated actively in the name of `tensor fossils' \cite{Jeong:2012df,Dai:2013kra,Brahma:2013rua,Dimastrogiovanni:2014ina}.
} As mentioned above, since there is similarity between Lorentz-violating massive gravity and
 supersolid inflation, it is expected that the quadrupolar modulation of the tensor power spectrum is also induced
 in the Lorentz-violating massive gravity.
 Therefore, we calculate the quadrupolar modulation of the tensor power spectrum induced by the tensor bispectrum
 and briefly discuss the detectability.%
 \footnote{For the discussion on the detection of primordial GWs with pulsar timing arrays like 
SKA~\cite{Janssen:2014dka}, see~\cite{Tsuneto:2018tif}.}

The rest of this paper is organized as follows. In section \ref{sec:dynamics_tensor_MTMG},
we briefly summarize the result on the tensor power spectrum in the scenario of ~\cite{FKMM}.  
Then, in section \ref{sec:tensor_bispectrum_MTMG}, we calculate the tensor bispectrum 
in the same scenario today based on the in-in formalism. In section \ref{sec:Modulation},
we calculate the quadrupolar modulation of the tensor power spectrum sourced by 
the tensor bispectrum and consider the detectability.
Section \ref{sec:conclusions} is devoted to conclusions and discussions.

\section{Dynamics of Massive Graviton}
\label{sec:dynamics_tensor_MTMG}

Here, we briefly explain our setup which was developed in Ref.~\cite{FKMM}. Our quadratic action for the tensor graviton $h_{ij}(\tau,\bm{x})$
is given by
\bea
S^{(2)}_h =\frac{\Mp^2}{8} \int \dd\tau \dd^3 x\, a^2 
\left[ h_{ij}' h_{ij}' -  \partial_l h_{ij}\partial_l h_{ij} -a^2 \mu^2 h_{ij} h_{ij} \right]\,,
\ena
where $M_{\rm Pl}$ is the reduced Planck mass, $a(\tau)$ is the scale factor, 
$\mu (\tau)$ is the mass of the tensor graviton, $\tau$ is the conformal time and a prime denotes its derivative, i.e. $X'\equiv \partial_\tau X$. The tensor gravitons can be decomposed and quantized as
\begin{equation}
h_{ij}(\tau,\bm{x})=\frac{2}{aM_{\rm Pl}}\sum_{\lambda=+,-}
\int \frac{\mathrm{d}^3 k}{(2\pi)^3} e^{i\bm{k\cdot x}}\,e_{ij}^\lambda(\hat{\bm k})  \left[v^\lambda_{k}(\tau)\hat{a}^\lambda_{\bm k}+{\rm h.c.} \right],
\label{hij def}
\end{equation}
where $e_{ij}^\lambda(\hat{\bm k})$ is the polarization tensor and $\hat{a}_{\bm k}/\hat{a}^\dag_{\bm k}$ are creation/annihilation operators satisfying the commutation relation,
$[\hat{a}^\lambda_{\bm k}, \hat{a}^{\dag\sigma}_{\bm p}]=(2\pi)^3\delta^{\lambda\sigma} \delta(\bm k-\bm p) $.%
\footnote{The polarization tensor $e_{ij}^\lambda(\hat{\bm k})$ generally satisfies the transverse-traceless condition $k_i   e_{ij}^\lambda(\hat{\bm k}) =  e_{ii}^\lambda(\hat{\bm k}) =0$.
Since we employ the circular polarization tensor, we also have
 the normalization conditions $ e_{ij}^{\lambda*}(\hat{\bm k}) e_{ij}^{\lambda'}(\hat{\bm k})  = \delta_{\lambda\lambda'}$  and the following properties $e_{ij}^{\lambda*}(\hat{\bm k})= e_{ij}^{-\lambda}(\hat{\bm k})  = e_{ij}^\lambda(-\hat{\bm k})$ for $\lambda=\pm$.}
Henceforth, we often suppress the polarization label $\lambda$ when it is not relevant.

To obtain the evolution of the mode function $v_k$, we need to specify $a(\tau), \mu(\tau)$ and the initial condition for $v_k(\tau)$.
For simplicity, we assume the de Sitter expansion $a \propto \tau^{-1}$ during inflation as well as instantaneous reheating followed by the radiation dominated era $a\propto\tau$, which gives
\begin{equation}
a(\tau)=\begin{cases}
-1/(H_{\inf}\tau) & (\tau<-\tau_r) \\
a_r\tau/\tau_r & (\tau>\tau_r) \\
\end{cases}.
\end{equation}
Here $H_{\inf}$ is the Hubble expansion rate during inflation and
$a_r$ is the scale factor at the reheating time $\tau_r= (a_r H_{\inf})^{-1}$.
Note that  in this treatment  
the conformal time $\tau$ jumps from $-1/(aH_{\inf})$ into $1/(aH_{\inf})$ at reheating for $a$ and $\mathrm{d} a/\mathrm{d} \tau$ to be continuous. We further assume a simple step-function behavior of the graviton mass,
\begin{align}
\mu(\tau)&=
\begin{cases} 
m & (\tau<\tau_m) \\
0 & (\tau>\tau_m) \\
\end{cases},
\label{m evolution}
\end{align}
where $\tau_m$ is a certain time during radiation dominated era.
Finally, we set the initial condition for the mode function to be that for the Bunch-Davies vacuum during inflation,
\begin{equation}
\lim_{k\tau\to -\infty}v_k(\tau) = \frac{1}{\sqrt{2 k}}e^{-i k\tau}.
\label{BD condition}
\end{equation}

Solving the equation of motion (EoM) based on the above setup, one can show that the mode function $v_k(\tau)$ during inflation is given by
\bea
v_k(\tau<\tau_r)= \frac{\sqrt{-\pi \tau}}{2} H_\nu^{(1)}(- k \tau)\,, \qquad
  \nu\equiv \sqrt{\frac{9}{4}-\frac{m^2}{H^2_{\inf}}}\,.
  \label{mode_function}
\ena
where $H_{\nu} ^{(1)}$ is the Hankel function of the first kind of order $\nu$. To discuss the mode function during inflation in more detail, it is useful to introduce a new dimensionless time variable $x\equiv -k\tau$.
Well after the horizon exit, $x \ll 1$, the mode function has the asymptotic form
\bea
\lim_{x \to 0} v_k (x) = \sqrt{\frac{\pi}{2k}}  
\left[\frac{1}{\Gamma(1+\nu)} \left(\frac{x}{2}\right)^{\frac12+\nu}
-i \frac{\Gamma (\nu)}{\pi} \left(\frac{x}{2}\right)^{\frac12-\nu} \right],
\quad \quad (\tau<\tau_r),
\label{mode_function_asym_large}
\ena
where $\Gamma (\nu)$ is the gamma function. 
Although the second term is always dominant in the magnitude, 
the first term which carries the real part of $v_k$ also plays an important role in the calculation of the tensor bispectrum, as we will see in the next section.

The ratio of the mode functions between the current massive case (${v_k^{\rm massive}(\tau)}$) and the usual massless case (${v_k^{\rm massless}(\tau)}$) is evaluated
at the end of inflation as
\begin{equation}
\Upsilon_k(\tau)\equiv \left|\frac{v_k^{\rm massive}(\tau)}{v_k^{\rm massless}(\tau)}\right| \quad\Longrightarrow\quad
\Upsilon_k(\tau_r)\simeq \frac{\Gamma(\nu)}{\Gamma(3/2)}\left|\frac{k\tau_r}{2}\right|^{\frac{3}{2}-\nu}.
\label{Upsilon def}
\end{equation}
Since $|k\tau_r|\ll1$ and $\nu<3/2$ for the massive graviton on super-horizon scales, the mode function is suppressed compared to the conventional massless case. 
On the other hand, after inflation ends and the Hubble expansion rate $H$ decreases, the super-horizon graviton modes behave as non-relativistic matter with $m>H>k/a$. Then the mode function is relatively amplified, $\Upsilon_k \propto a^{1/2}$, compared to the massless graviton modes which behave as radiation. This amplification continues until the graviton mass vanishes at $\tau=\tau_m$.
The ratio of the mode functions is then given by~\cite{FKMM}
\begin{equation}
\Upsilon_k(\tau\gg \tau_m)=\gamma_k \sqrt{\frac{\tau_m}{\tau_r}}\,
\frac{\Gamma(\nu)}{\Gamma(3/2)}\left|\frac{k\tau_r}{2}\right|^{\frac{3}{2}-\nu},
\label{late time Upsilon}
\end{equation}
where $\gamma_k$ has a rather lengthy expression obtained by solving the junction conditions, but it is $\mathcal{O}(1)$ for $m/H_{\inf}\sim 1$.
As a result, the dimensionless power spectrum of the tensor modes
is written as~\cite{FKMM}
\bea
\mcP_h(k,\tau\gg\tau_m) =\tilde{\gamma}_k^2\ |k\tau_r|^{3-2\nu}\frac{\tau_m}{\tau_r}
\ \mcP_h^{\rm massless}(k,\tau),
\label{tensor_powerspectrum}
\ena
with $\tilde{\gamma}_k\equiv \gamma_k \times 2^{\nu-3/2}\Gamma(\nu)/\Gamma(3/2)$.
Here, $\mcP_h^{\rm massless}$ denotes the usual power spectrum of the massless tensor modes from inflation,
\begin{equation}
\mcP_h^{\rm massless}(k,\tau)=\mcT_k^2(\tau)\,\frac{2H_{\inf}^2}{\pi^2\Mpl^2},
\end{equation}
with the transfer function of the massless tensor modes, 
\begin{equation}
\mcT_k(\tau)\equiv \frac{h_k^{\rm massless}(\tau)}{h_k^{\rm massless}(\tau_r)}
=\frac{a(\tau_r)}{a(\tau)}\frac{v_k^{\rm massless}(\tau)}{v_k^{\rm massless}(\tau_r)}.
\label{transfer def}
\end{equation}
If gravitons keep the mass for a while after inflation $\tau_m\gg \tau_r$, it overcomes the dumping factor in Eq.~\eqref{Upsilon def} and the gravitational waves  can be substantially amplified for relevant modes with $|k\tau_r|^{3-2\nu}\tau_m/\tau_r \gg 1$.
The tensor tilt is
\bea
n_T\equiv \frac{\dd\ln \mcP_h(\tau_r)}{\dd \ln k}=3-2\nu,
\label{tilt}
\ena
where the $\mathcal{O}(\epsilon_H)$ slow-roll correction is ignored
under the assumption $m/H_{\inf}=\mathcal{O}(1)$. 
Notice that the tilt of the power spectrum is blue, $n_T>0$, because
the mode function decays on super-horizon scales during inflation.

\section{Tensor Bispectrum}
\label{sec:tensor_bispectrum_MTMG}

In this section, we calculate the tensor bispectrum. We consider the following interaction Hamiltonian at the third order for the tensor perturbation:
\begin{equation}
H_{\rm int} = H_{\rm int}^{\rm (GR)} + H_{\rm int}^{\rm (mass)}\,,
\end{equation}
with
\begin{align}
H_{\rm int}^{\rm (GR)}&=-\frac{\Mp^2}{4}  a^2 \int \dd^3 x\, h_{ij} h_{kl} \left(\partial_j\partial_l h_{ik} -\partial_i\partial_j \frac12 h_{kl} \right)\,,
\label{H GR}
\\
H_{\rm int}^{\rm (mass)}&=-g\frac{\Mp^2}{4}  a^4 \int \dd^3 x\, h_{ij} h_{jk} h_{ki}\,,
\label{H mass}
\end{align}
where the coefficient $g(\tau)$ depends on time.
$H_{\rm int} ^{\rm (GR)}$ is found even in general relativity (GR) around de Sitter space.
On the other hand, $H_{\rm int}^{\rm (mass)}$ is absent in GR and 
it arises from the same argument as the graviton mass term as 
discussed in appendix~\ref{appendix_MTMG}.
We assume $g$ is also constant during inflation in the same way as the graviton mass $\mu$ (see eq.~\eqref{m evolution}).
It is worth mentioning that
$H_{\rm int}^{\rm (mass)}$ also appears in supersolid inflation~\cite{Ricciardone:2016lym,Ricciardone:2017kre}, which suggests that the appearance of this interaction is a generic feature of
the breaking of the space reparameterization symmetry.

The three-point function for the tensor mode $h_{ij}$ can be calculated based on the in-in formalism
\cite{Maldacena:2002vr,Weinberg:2005vy},
\begin{align}
&\langle h_{i_1 j_1}(\tau,\bm{k}_1) h_{i_2 j_2}(\tau,\bm{k}_2)h_{i_3 j_3}(\tau,\bm{k}_3)\rangle
=i \int^\tau_{-\infty} \dd \eta \langle [ H_{\rm int}(\eta),  h_{i_1 j_1}(\tau,\bm{k}_1) h_{i_2 j_2}(\tau,\bm{k}_2)h_{i_3 j_3}(\tau,\bm{k}_3)]\rangle.
\label{in_in_formalism}
\end{align}
From this three-point function, we define un-contracted bispectrum for later convenience as
\begin{align}
\langle h_{i_1 j_1}(\tau,\bm{k}_1) h_{i_2 j_2}(\tau,\bm{k}_2)h_{i_3 j_3}(\tau,\bm{k}_3)\rangle
\equiv (2\pi)^3 \delta(\bm k_1+\bm k_2+\bm k_3)B_{i_1 j_1 i_2 j_2 i_3 j_3}(k_1,k_2,k_3),
\label{uncBdef}
\end{align}
where $k_i \equiv |\bm{k}_i|$. Taking the following sum over the indices, one can compute the tensor bispectrum from it,
\begin{equation}
B_h(k_1,k_2,k_3)=\delta_{j_1 i_2}\delta_{j_2 i_3}\delta_{j_3 i_1}
B_{i_1 j_1 i_2 j_2 i_3 j_3}(k_1,k_2,k_3).
\end{equation}
The un-contracted bispectrum $B_{i_1 j_1 i_2 j_2 i_3 j_3}$ can be split into two parts which are contributed by $H_{\rm int}^{\rm (GR)}$ and $H_{\rm int}^{\rm (mass)}$, respectively, as
\begin{equation}
B_{i_1 j_1 i_2 j_2 i_3 j_3}= B_{i_1 j_1 i_2 j_2 i_3 j_3}^{\rm (GR)}+B_{i_1 j_1 i_2 j_2 i_3 j_3}^{\rm (mass)}.
\end{equation}
Plugging eqs.~\eqref{H GR} and \eqref{H mass} into \eqref{in_in_formalism},
we obtain
\begin{align}
B_{i_1 j_1 i_2 j_2 i_3 j_3}^{\rm (GR)} &=-\frac{32}{a^3\Mpl^4}\mathcal{I}^{\rm (GR)}(k_1,k_2,k_3;\tau)
\,\mathcal{E}^{\rm (GR)}_{i_1j_1i_2j_2i_3j_3}(\hat{\bm k}_1,\hat{\bm k}_2,\hat{\bm k}_3)\,,
\label{B GR def}
\\
B_{i_1 j_1 i_2 j_2 i_3 j_3}^{\rm (mass)} &= 
\frac{192}{a^3\Mpl^4}\mathcal{I}^{\rm (mass)}(k_1,k_2,k_3;\tau)
\,\mathcal{E}^{\rm (mass)}_{i_1j_1i_2j_2i_3j_3}(\hat{\bm k}_1,\hat{\bm k}_2,\hat{\bm k}_3),
\label{B mass def}
\end{align}
where $\mathcal{I}^{(X)}$ are basically time integrals of the mode functions,
\begin{align}
\mathcal{I}^{\rm (GR)}(k_1,k_2,k_3;\tau)&\equiv k_1^2\int^\tau_{-\infty} \dd \eta\, a^{-1}(\eta) \, {\rm Im}\left[v_{k_1}^*(\tau)v_{k_2}^*(\tau)v_{k_3}^*(\tau)
v_{k_1}(\eta)v_{k_2}(\eta)v_{k_3}(\eta)\right],
\label{IGR def}
\\
\mathcal{I}^{\rm (mass)}(k_1,k_2,k_3;\tau)&\equiv\int^\tau_{-\infty} \dd \eta\,  a(\eta) g(\eta)\, {\rm Im}\left[v_{k_1}^*(\tau)v_{k_2}^*(\tau)v_{k_3}^*(\tau)
v_{k_1}(\eta)v_{k_2}(\eta)v_{k_3}(\eta)\right],
\label{Imass def}
\end{align}
while $\mathcal{E}^{(X)}$ are combinations of the tensor polarizations,
\begin{align}
\mathcal{E}^{\rm (GR)} _{i_1j_1i_2j_2i_3j_3}(\hat{\bm k}_1,\hat{\bm k}_2,\hat{\bm k}_3)=&
\bigg[\Pi_{i_1 j_1,ij}(\hat{\bm k}_1)\Pi_{i_2 j_2,kl}(\hat{\bm k}_2)
\left(\kappa_{3 j}\kappa_{3l}\Pi_{i_3 j_3,i k}(\hat{\bm k}_3)-\frac{1}{2}\kappa_{3i}\kappa_{3j}\Pi_{i_3 j_3,kl}(\hat{\bm k}_3)\right)
\notag\\&~~~~~~~~~~~~~~~~~~~~~~~~~+{\rm 5\ permutation\ terms \ w.r.t. \  1,2,3}\bigg],
\\
\mathcal{E}^{\rm (mass)}_{i_1j_1i_2j_2i_3j_3}(\hat{\bm k}_1,\hat{\bm k}_2,\hat{\bm k}_3)=&\Pi_{i_1 j_1,ij}(\hat{\bm k}_1)\Pi_{i_2 j_2,jk}(\hat{\bm k}_2)\Pi_{i_3 j_3,ki}(\hat{\bm k}_3)
\end{align}
with
\begin{equation}
\bm{\kappa}_i\equiv \frac{\bm{k}_i}{k_1}\;\;(i=2,\;3)\;,
\qquad
\Pi_{ij,kl}(\hat{\bm k})\equiv \sum_{\lambda}
e_{ij}^{\lambda}(\hat{\bm k})
e_{kl}^{*\lambda}(\hat{\bm k}).
\end{equation}
We shall also adopt the notation $\kappa_i \equiv |\bm{\kappa}_i|=k_i/k_1$. 

In what follows in this section, we will evaluate  $\mathcal{I}^{(X)}$.
As discussed in appendix~\ref{Real Part versus Imaginary Part}, it can be shown that $\mathcal{I}^{(X)}$
are well approximated by
\begin{align}
\mathcal{I}^{\rm (GR)}(\tau)&\simeq
{\rm Im}\left[v_{k_1}^*(\tau)v_{k_2}^*(\tau)v_{k_3}^*(\tau)\right]
k_{1}^2\int^\tau_{-\infty} \dd \eta\, a^{-1}(\eta) \, 
{\rm Re}\left[v_{k_1}(\eta)v_{k_2}(\eta)v_{k_3}(\eta)\right],
\label{I GR}
\\
\mathcal{I}^{\rm (mass)}(\tau)&\simeq
{\rm Im}\left[v_{k_1}^*(\tau)v_{k_2}^*(\tau)v_{k_3}^*(\tau)\right]
\int^\tau_{-\infty} \dd \eta\,  a(\eta) g(\eta)\, 
{\rm Re}\left[v_{k_1}(\eta)v_{k_2}(\eta)v_{k_3}(\eta)\right].
\label{I mass}
\end{align}
We will evaluate these time integrals and the prefactor ${\rm Im}[v_{k_1}^*v_{k_2}^*v_{k_3}^*]$ in Sec.~\ref{Evaluating time integral during inflation} and \ref{Evolution after inflation}, respectively.

\subsection{Evaluating time integral during inflation}
\label{Evaluating time integral during inflation}

\begin{figure}[tbp]
  \begin{center}
  \includegraphics[width=100mm]{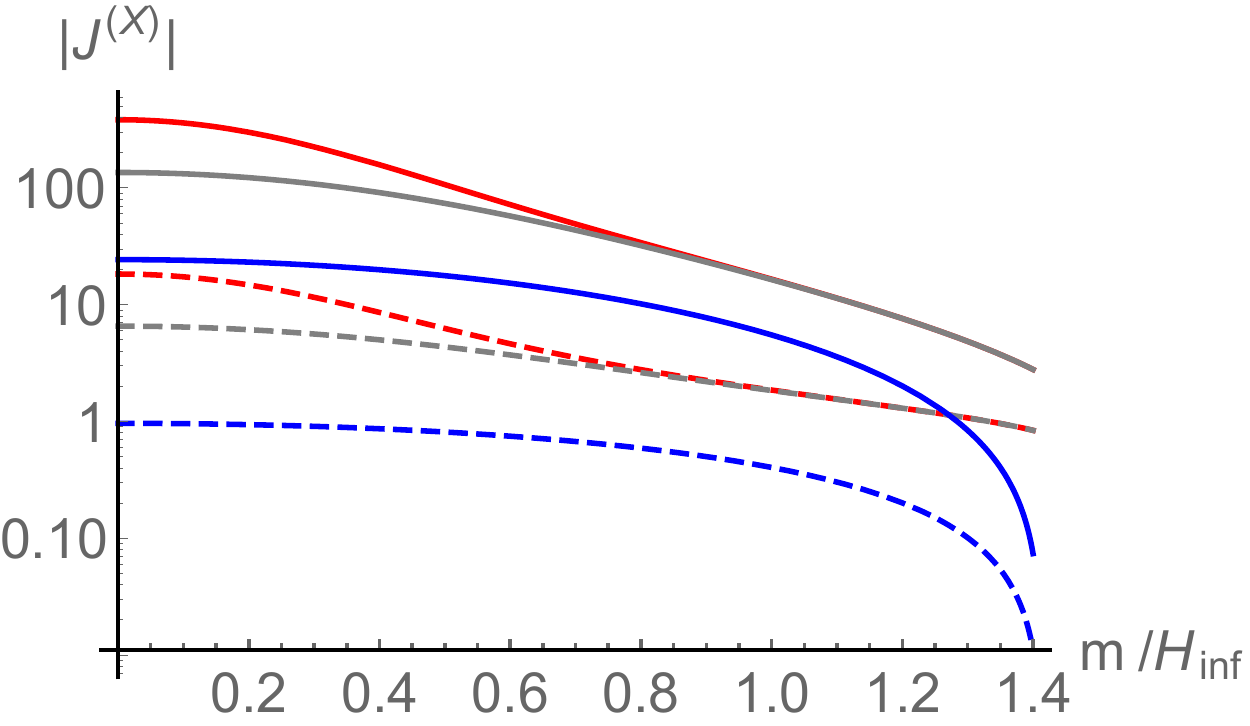}
  \end{center}
  \caption{The numerical results of $|\mathcal{J}^{\rm (X)}(\kappa_2,\kappa_3)|$ defined in Eq.~\eqref{J def} with $X=\rm GR$ (blue) and $X=\rm mass$ (red and grey)
are shown as functions of $m/H_{\inf}$. The solid and dashed line denote the squeezed configuration $(\kappa_2=1, \kappa_3=0.1)$ and the equilateral configuration $(\kappa_2=\kappa_3=1)$, respectively. The lower bound of the integral range of $|\mathcal{J}^{\rm (mass)}|$ is $|k\tau_r|=10^{-15}$ (red) and $10^{-5}$ (grey), while $|\mathcal{J}^{\rm (GR)}|$ (blue) does not depend on it.  $|\mathcal{J}^{\rm (mass)}|$ is amplified for $m/H_{\inf}\lesssim 0.7$ as $|k\tau_r|$ decreases. }
 \label{Fig_Jm}
\end{figure}

Concentrating on the inflationary era,  $\tau<\tau_r$, 
one can rewrite the time integrals in Eqs.~\eqref{I GR} and \eqref{I mass} as
\begin{equation}
\int^{\tau_r}_{-\infty} \dd \eta\, a^{\mp 1}(\eta) \, 
{\rm Re}\left[v_{k_1}(\eta)v_{k_2}(\eta)v_{k_3}(\eta)\right]
=\frac{\pi^{3/2}}{8k^{5/2}_1}\left(\frac{H_{\inf}}{k_1}\right)^{\pm1}
\mathcal{J}^{\rm (GR/mass)},
\label{Re integ}
\end{equation}
with dimensionless integrals
\begin{equation}
\mathcal{J}^{\rm (GR/mass)}(\kappa_2,\kappa_3)\equiv
\int^{\infty}_{|k_1 \tau_r|} \dd y \,
y^{\frac{3}{2}\pm 1}\, {\rm Re}
\left[ H_\nu^{(1)}(y)H_\nu^{(1)}(\kappa_2 y)H_\nu^{(1)}(\kappa_3 y)\right],
\label{J def}
\end{equation}
where $g(\eta)$ is assumed to be constant during inflation and  $y\equiv -k_1 \eta$ is introduced as a new dummy variable.
$\mathcal{J}^{\rm (GR/mass)}$ depends on $\kappa_2,\kappa_3$ and $\nu=\sqrt{9/4-m^2/H^2_{\inf}}$
and should be numerically evaluated.
In Fig.~\ref{Fig_Jm}, we show their dependence on $m/H_{\inf}$ for the equilateral ($\kappa_2=\kappa_3=1$) and the squeezed ($\kappa_3\ll\kappa_2=1$) configurations.
The following three observations are found in Fig.~\ref{Fig_Jm}:
(i) $|\mathcal{J}^{\rm (mass)}|$ is always larger than $|\mathcal{J}^{\rm (GR)}|$.
(ii) The squeezed configuration (solid) is larger than the equilateral configuration (dashed).
(iii) $|\mathcal{J}^{\rm (mass)}|$ increases  as $|k\tau_r|$ decreases for $|k\tau_r|\lesssim0.7$, while $|\mathcal{J}^{\rm (GR)}|$ is insensitive to $|k\tau_r|\ll1$.
%
%

Comparing $\mathcal{I}^{\rm (mass)}$ and $\mathcal{I}^{\rm (GR)}$, we find
\begin{equation}
\frac{\mathcal{I}^{\rm (mass)}}{\mathcal{I}^{\rm (GR)}}=
\frac{g_{\inf}}{H_{\inf}^2}\,
\frac{\mathcal{J}^{\rm (mass)}}{\mathcal{J}^{\rm (GR)}},
\label{I ratio}
\end{equation}
where $g_{\inf}$ is the value of $g$ during inflation.
Since $\mathcal{J}^{\rm (mass)}$ is larger than $\mathcal{J}^{\rm (GR)}$
by almost an order of magnitude as seen in Fig.~\ref{Fig_Jm},
we expect that the contribution to the tensor bispectrum is dominated
by $B_{i_1 j_1 i_2 j_2 i_3 j_3}^{\rm (mass)}$ rather than 
$B_{i_1 j_1 i_2 j_2 i_3 j_3}^{\rm (GR)}$, if $g_{\inf}\gtrsim H_{\inf}^2$.
We will explicitly confirm this expectation with numerical computations in the next section.

One can also see in Fig.~\ref{Fig_Jm} that the squeezed configuration case is much larger than the equilateral case.
We find that $\mathcal{J}^{\rm (GR/mass)}$ diverges
in the squeezed limit as
\begin{align}
\lim_{\kappa_2\to1, \kappa_3\to0} \mathcal{J}^{\rm (GR/mass)}
&\simeq
-\lim_{\kappa_3\to0}\int^{\infty}_{|k_1 \tau_r|} \dd y \,
y^{\frac{3}{2}\pm 1}\, {\rm Im}
\left[ \left(H_\nu^{(1)}(y)\right)^2\right]
{\rm Im}\left[H_\nu^{(1)}(\kappa_3 y)\right]
\notag\\
&=\lim_{\kappa_3\to0}\kappa_3^{-\nu}\times\left( 
\frac{2^\nu\Gamma(\nu)}{\pi}\,
\int^{\infty}_{|k_1 \tau_r|} \dd y \,
y^{\frac{3}{2}-\nu\pm 1}\, {\rm Im}
\left[ \left(H_\nu^{(1)}(y)\right)^2\right]\right).
\label{squeezed J}
\end{align}
This implies that the tensor bispectrum peaks at the squeezed configuration, and the degree of divergence in the squeezed limit is solely determined by $\nu=\sqrt{9/4-m^2/H^2_{\inf}}$.

Although $\mathcal{J}^{\rm (GR)}$ is not sensitive to the lower limit of the integral, $|k_1 \tau_r|\ll1$, we found $\mathcal{J}^{\rm (mass)}$
depends on it for $m/H_{\inf}\lesssim 0.7$.
Well after all the modes with $k_1, k_2, k_3$ exit the horizon, the integrand of $\mathcal{J}^{\rm (GR/mass)}$ in Eq.~\eqref{J def} evolve as,
\begin{equation}
 \,
y^{\frac{3}{2}\pm 1}\, {\rm Re}
\left[ H_\nu^{(1)}(y)H_\nu^{(1)}(\kappa_2 y)H_\nu^{(1)}(\kappa_3 y)\right]
\ \sim\  
y^{\frac{3}{2}-\nu\pm 1}.
\end{equation}
Hence, the massless limit ($\nu=3/2$) of $\mathcal{J}^{\rm (mass)}$
exhibits a logarithmic enhancement. Indeed, we can explicitly calculate 
$\mathcal{J}^{\rm (mass)}$ in the massless limit as
\begin{equation}
\lim_{m\to0}\mathcal{J}^{\rm (mass)}=
\left(1+\kappa_2^3+\kappa_3^3\right) (N_{{k}_1}-\gamma_E)
+\frac{1}{3} (1+\kappa_2+\kappa_3) \left(4+4\kappa_2^2+4\kappa_3^2-\kappa_2-\kappa_3-\kappa_2 \kappa_3\right),
\end{equation}
where $N_{{k}_1}\equiv -\log|k_1\tau_r|$ and $\gamma_E$ is Euler's constant.\footnote{Basically this integral was
calculated in Ref.~\cite{Ricciardone:2016lym} (see  Eq.~(72)  of the paper).
There are, however, a couple of typos in the paper including the absence of the term proportional to $N_{{k}_1}$.
For these points, we contacted the authors of Ref.~\cite{Ricciardone:2016lym} and we have got the agreement.}
Nevertheless, in this paper, we focus on the cases of $m/H_{\inf}=\mathcal{O}(1)$
with which we obtain the amplified blue-tilted GW power spectrum, Eq.~\eqref{tensor_powerspectrum}.
Then, not only $\mathcal{J}^{\rm (GR)}$ but also $\mathcal{J}^{\rm (mass)}$
becomes constant for a sufficiently small $|k_1\tau_r|$.
Therefore, we can ignore the late time contribution to $\mathcal{J}^{\rm (GR/mass)}$.

\subsection{Evolution after inflation}
\label{Evolution after inflation}

Now we consider the factor ${\rm Im}\left[v_{k_1}^*(\tau)v_{k_2}^*(\tau)v_{k_3}^*(\tau)\right]$
in Eqs.~\eqref{I GR} and \eqref{I mass} which represents the post-inflationary evolution of the bispectrum.
Evaluating at the end of inflation and using eq.~\eqref{mode_function_asym_large} again, one finds
\begin{equation}
a^{-3}(\tau_r)\,
{\rm Im}\left[v_{k_1}^*(\tau_r)v_{k_2}^*(\tau_r)v_{k_3}^*(\tau_r)\right]
\simeq
\frac{2^{3\nu-3}\Gamma^3(\nu)H_{\inf}^3}{\pi^{3/2}\kappa_2^\nu \kappa_3^\nu k_1^{9/2}}
\,|k_1\tau_r|^{\frac{3}{2}(3-2\nu)},
\end{equation}
where $a^{-3}$ was multiplied because the bispectrum has this prefactor in Eqs.~\eqref{B GR def} and \eqref{B mass def}
which originally comes from the fact, $h(\tau)\propto v(\tau)/a(\tau)$.
The evolution after inflation is given by a factor from Eqs.~\eqref{late time Upsilon} and \eqref{transfer def} 
\begin{equation}
\frac{v_k(\tau)}{v_k(\tau_r)}= \sqrt{\frac{\tau_m}{\tau_r}}\,\gamma_k\,
\frac{a(\tau)}{a(\tau_r)}\mcT_k(\tau).
\end{equation}
Now we obtain
\begin{equation}
a^{-3}(\tau)\,
{\rm Im}\left[v_{k_1}^*(\tau)v_{k_2}^*(\tau)v_{k_3}^*(\tau)\right]
\simeq H_{\inf}^3
\frac{\mcT_{k_1}(\tau)\mcT_{k_2}(\tau)\mcT_{k_3}(\tau)}{2^{3/2}\kappa_2^\nu \kappa_3^\nu k_1^{9/2}}
\tilde{\gamma}_{k_1}\tilde{\gamma}_{k_2}\tilde{\gamma}_{k_3}
\left[\,|k_1\tau_r|^{(3-2\nu)}\frac{\tau_m}{\tau_r}\right]^{\frac{3}{2}}.
\label{3/2 factor}
\end{equation}
Here, the last factor $\tilde{\gamma}_{k_1}\tilde{\gamma}_{k_2}\tilde{\gamma}_{k_3}[|k_1\tau_r|^{(3-2\nu)}\tau_m/\tau_r]^{3/2}$ reminds us of the expression for the power spectrum, Eq.~\eqref{tensor_powerspectrum}.
Since the power spectrum ($\propto h^2$) gains the  factor $\tilde{\gamma}_k^2\,|k_1\tau_r|^{(3-2\nu)}\tau_m/\tau_r$, it is reasonable for the bispectrum ($\propto h^3$) to acquire it to the $3/2$ power.

In the case of the instant reheating, the transfer function for the
massless tensor at the present time $\mcT_k(\tau_0)$
is given by~\cite{Turner:1993vb}
\begin{equation}
\mcT_k(\tau_0)=\Omega_{m0}\sqrt{\frac{g_*(T_{\rm in})}{g_{*0}}}
\left(\frac{g_{*s0}}{g_{*s}(T_{\rm in})}\right)^{2/3}
\frac{3 j_1(k\tau_0)}{k\tau_0} \tilde{T}_1(k/k_{eq}),
\end{equation}
where $j_1(x)=(\sin (x)/x-\cos (x))/x$, $\tilde{T}_1(x)=1+1.57x+3.42x^2$,
$T_{\rm in}$ is the temperature of the universe when the mode reenters the horizon and $k_{\rm eq}=7.1\times10^{-2}\Omega_{m0}h^2 {\rm Mpc}^{-1}$ is
the wavenumber corresponding to the horizon scale of the matter-radiation equality.

\subsection{Tensor bispectrum today}

Putting the results of the previous subsections altogether, we obtain
the two contributions to the contracted bispectrum at the present time as
\begin{align}
k_1^2 k_2^2 k_3^2\,B_{h}^{\rm (GR)} &=-\frac12 (2 \pi)^{3/2} \frac{H_{\inf}^4}{\Mpl^4}
\frac{\mathcal{E}^{\rm (GR)} \mathcal{J}^{\rm (GR)}}{\kappa_2^{\nu-2} \kappa_3^{\nu-2}}
\mcT_{k_1}\mcT_{k_2}\mcT_{k_3}(\tau_0)
\tilde{\gamma}_{k_1}\tilde{\gamma}_{k_2}\tilde{\gamma}_{k_3}
\left[|k_1\tau_r|^{(3-2\nu)}\frac{\tau_m}{\tau_r}\right]^{\frac{3}{2}},
\label{B GR today}
\\
k_1^2 k_2^2 k_3^2\,B_{h}^{\rm (mass)} &=3(2\pi)^{3/2}\,\frac{g_{\inf}H_{\inf}^2}{\Mpl^4}
\frac{\mathcal{E}^{\rm (mass)}\mathcal{J}^{\rm (mass)}}{\kappa_2^{\nu-2} \kappa_3^{\nu-2}}
\mcT_{k_1}\mcT_{k_2}\mcT_{k_3}(\tau_0)
\tilde{\gamma}_{k_1}\tilde{\gamma}_{k_2}\tilde{\gamma}_{k_3}
\left[|k_1\tau_r|^{(3-2\nu)}\frac{\tau_m}{\tau_r}\right]^{\frac{3}{2}},
\label{B mass today}
\end{align}
where  we multiplied $B_h^{\rm (GR/mass)}$ by $(k_1 k_2 k_3)^2$ in order to make them dimensionless. 
Here, the contracted polarization tensors
$\mathcal{E}^{\rm (GR/mass)}\equiv \delta_{j_1 i_2}\delta_{j_2 i_3}\delta_{j_3 i_1}\mathcal{E}^{\rm (GR/mass)}_{i_1 j_1 i_2 j_2 i_3 j_3}$
are computed as 
\begin{align}
\mathcal{E}^{\rm (GR)}&=
\frac{1}{1024 \kappa_2^4 \kappa_3^4}\left(\kappa_2^4-2 \kappa_2^2 \left(\kappa_3^2+1\right)+\left(\kappa_3^2-1\right)^2\right)^2 \bigg[ \kappa_2^6+15 \kappa_2^4 \left(\kappa_3^2+1\right)
\notag\\&\qquad\qquad\qquad\qquad
+15 \kappa_2^2 \left(\kappa_3^4+6 \kappa_3^2+1\right)
+\left(\kappa_3^6+15 \kappa_3^4+15\kappa_3^2+1\right)\bigg],
\\
\mathcal{E}^{\rm (mass)}&=\frac{1}{512 \kappa_2^4 \kappa_3^4}
\left(-2 \left(\kappa_2^2+1\right) \kappa_3^2+\left(\kappa_2^2-1\right)^2+\kappa_3^4\right)^2 \left(\kappa_2^4+6 \kappa_2^2 \left(\kappa_3^2+1\right)+\kappa_3^4+6 \kappa_3^2+1\right).
\end{align}
For instance, their values at the squeezed and equilateral configurations are
\begin{align}
&\mathcal{E}^{\rm (GR)}=\frac{1}{2}, 
& 
&\mathcal{E}^{\rm (mass)}=\frac{1}{4}. 
& 
&({\rm Squeezed:}\ \ \kappa_2=1, \kappa_3\to 0)
\\
&\mathcal{E}^{\rm (GR)}=\frac{1647}{1024}, 
& 
&\mathcal{E}^{\rm (mass)}=\frac{189}{512}. 
& 
&({\rm Equilateral:}\ \ \kappa_2=\kappa_3=1)
\end{align}
%
\begin{figure}[tbp]
    \hspace{-2mm}
  \includegraphics[width=70mm]{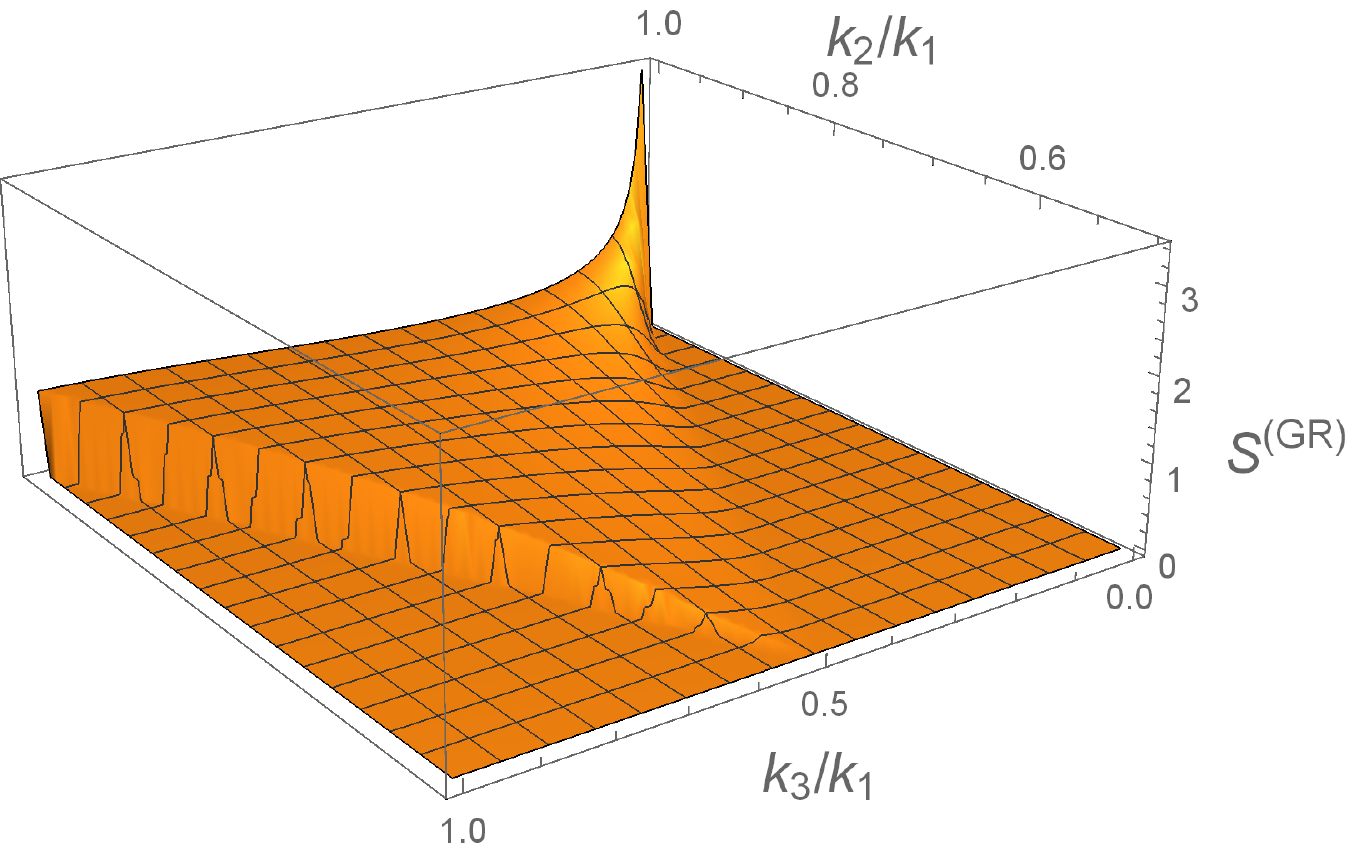}
  \hspace{5mm}
  \includegraphics[width=80mm]{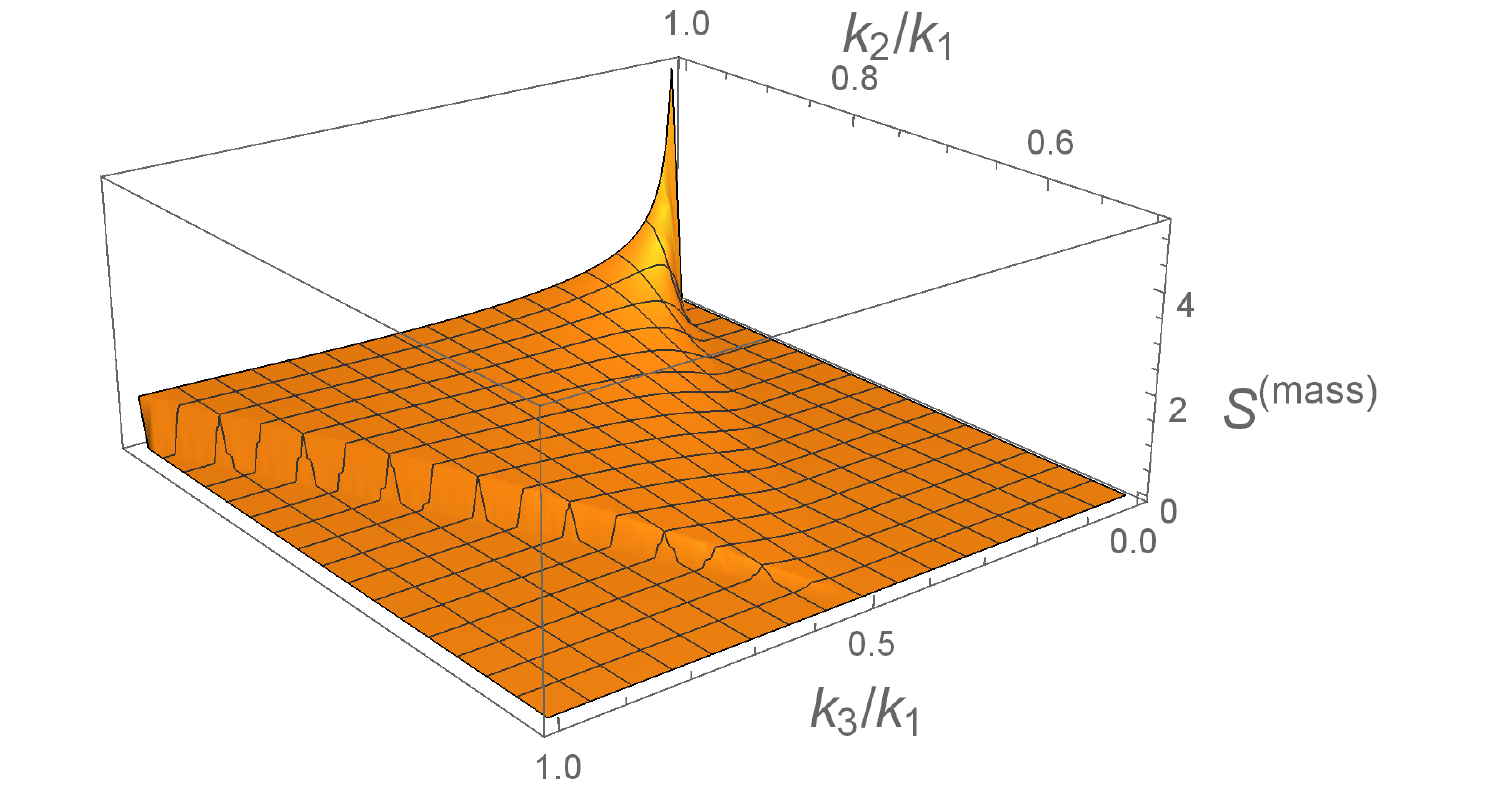}
  \caption
 {The shape functions $S^{\rm (GR)}$ (left panel) and $S^{\rm (mass)}$ (right panel) defined in Eq.~\eqref{shape def} are shown for $m/H_{\inf}=0.8$.
Both of them diverge at the squeezed limit, $\kappa_2\to1$ and $\kappa_3\to 0$.}
 \label{shape functions}
\end{figure}
%
We can also define their shape functions as
\begin{align}
S^{\rm (X)}(\kappa_2,\kappa_3)=\mathcal{N}^{\rm (X)}
\kappa_2^{2-\nu}\kappa_3^{2-\nu}
\mathcal{E}^{\rm (X)} (\kappa_2,\kappa_3)\mathcal{J}^{\rm (X)}(\kappa_2,\kappa_3),
\qquad ({\rm X= GR/mass})
\label{shape def}
\end{align}
where the factor $\mathcal{N}^{\rm (X)}=\left[\mathcal{E}^{\rm (X)}\mathcal{J}^{\rm (X)}(\kappa_2=\kappa_3=1)\right]^{-1}$ normalizes the shape function at the equilateral configuration such that $S^{(X)}(\kappa_2=\kappa_3=1)=1$.
From Eq.~\eqref{squeezed J}, one finds that the asymptotic behavior of the shape functions in the squeezed limit is the same for the two contributions,
\begin{equation}
\lim_{\kappa_2\to 1, \kappa_3\to 0}S^{\rm (X)}(\kappa_2,\kappa_3)\propto \kappa_3^{2-2\nu},
\qquad ({\rm X= GR/mass})
\end{equation}
and $(k_1k_2k_3)^2B_h$ diverges in the squeezed limit for $\nu>1\Longleftrightarrow m/H_{\inf}<\sqrt{5}/2\approx 1.12$.

\begin{figure}[tbp]
  \begin{center}
  \includegraphics[width=100mm]{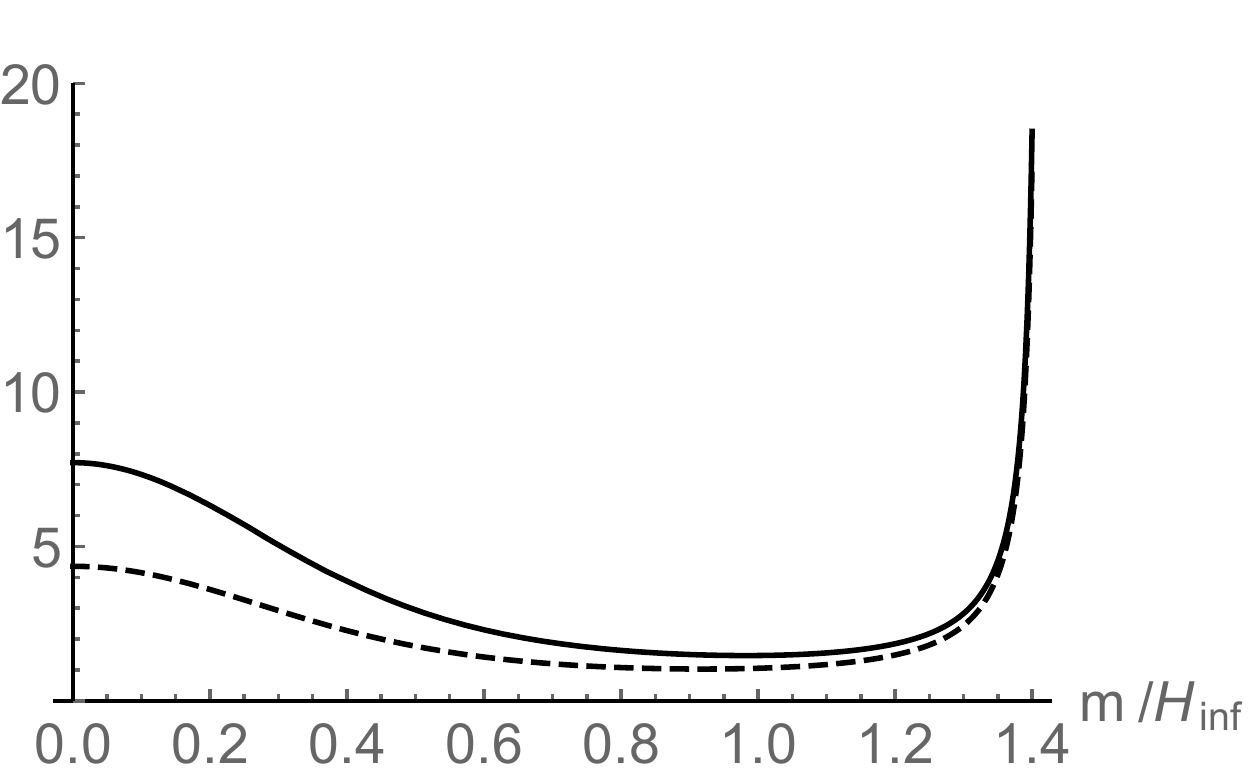}
  \end{center}
  \caption{$|\mathcal{E}^{\rm (mass)}\mathcal{J}^{\rm (mass)}/\mathcal{E}^{\rm (GR)}\mathcal{J}^{\rm (GR)}|$ which appears in the ratio $|B_{h}^{\rm (mass)}/B_{h}^{\rm (GR)}|$ in  Eq.~\eqref{B ratio} is plotted for $0\le m/H_{\inf}\le 1.4$.
The solid and dashed line denote the squeezed configuration $(\kappa_2=1, \kappa_3=0.1)$ and the equilateral configuration $(\kappa_2=\kappa_3=1)$, respectively. This plot implies that $|\mathcal{E}^{\rm (mass)}\mathcal{J}^{\rm (mass)}/\mathcal{E}^{\rm (GR)}\mathcal{J}^{\rm (GR)}|$ is always larger than unity. }
 \label{Fig_ratio}
\end{figure}

Let us compare the amplitudes of the two contributions,
$B_h^{(\rm mass)}$ and $B_h^{(\rm GR)}$.
Their ratio  is given by
\begin{equation}
\left|\frac{B_{h}^{\rm (mass)}}{B_{h}^{\rm (GR)}}\right|=
6\frac{g_{\inf}}{H_{\inf}^2}
\, 
\left|\frac{\mathcal{E}^{\rm (mass)}\mathcal{J}^{\rm (mass)}}{\mathcal{E}^{\rm (GR)}\mathcal{J}^{\rm (GR)}}\right|
\gtrsim 6\frac{g_{\inf}}{H_{\inf}^2}.
\label{B ratio}
\end{equation}
Here, $|\mathcal{E}^{\rm (mass)}\mathcal{J}^{\rm (mass)}/\mathcal{E}^{\rm (GR)}\mathcal{J}^{\rm (GR)}|$ is plotted in Fig.~\ref{Fig_ratio}
and is shown to be larger than unity irrespective of $m/H_{\inf}$ or configuration.
%
%
Therefore, for $g_{\inf}/H_{\inf}^2 = \mathcal{O}(1)$, the new contribution to the bispectrum $B_h^{(\rm mass)}$  dominates the conventional one $B_h^{\rm (GR)}$.

It is interesting to note that the bispectrum divided by the square root of the power spectra can be written in a simple expression,
\begin{align}
\frac{(k_1k_2k_3)^2B_h(k_1,k_2,k_3)}{[\mcP_h(k_1)\mcP_h(k_2)\mcP_h(k_3)]^{1/2}}
&\simeq 
\frac{3\pi^{9/2}g_{\inf}}{H_{\inf}\Mpl}\mathcal{E}^{\rm (mass)}\mathcal{J}^{\rm (mass)}
(\kappa_2 \kappa_3)^{\frac{7}{2}-2\nu},
\end{align}
where the sub-leading contribution from $B_h^{\rm (GR)}$ is ignored.
Here, the common factor discussed below Eq.~\eqref{3/2 factor}
which enhances both the  power spectrum and bispectrum is cancelled out.

\section{Quadrupolar modulation of tensor power spectrum}
\label{sec:Modulation}

In this section we discuss the detectability of the primordial tensor bispectrum calculated in the previous section.
Although it seems that the simplest way to detect the tensor bispectrum is the direct measurement
 of it at interferometer scales,  it was shown that
 the bispectrum cannot be probed directly ~\cite{Bartolo:2018evs,Bartolo:2018rku}.
 Instead of this, 
 we briefly discuss the detectability based on the modulation of the tensor power spectrum
induced by the squeezed tensor bispectrum.

The tensor power spectrum with polarization $\lambda_1$
modulated by a long tensor mode with polarization $\lambda_3$ and wavevector $\bm{k}_3$
is given by
\bea
\lim_{k_3 \to 0} \langle  h^{\lambda_1}_{\bm{k}_1} h^{\lambda_1}_{\bm{k}_2}  \rangle_{h_{\bm{k}_3}^{s_3}} ^\prime
\simeq \langle h^{\lambda_1}_{\bm{k}_1} h^{\lambda_1}_{\bm{k}_2} \rangle ^\prime
+ \lim_{k_3 \to 0} \left(  h_{\bm{k}_3}^{\lambda_3} \frac{\langle  h^{\lambda_1}_{\bm{k}_1}h^{\lambda_1}_{\bm{k}_2} h_{\bm{k}_3}^{\lambda_3}  \rangle^{\prime}}
{P_h ^{\lambda_3} (k_3)} \right)\,,
\label{expression_powerspectrum_tensor_ipoltensor}
\ena
where the prime on the expectation value $\langle \cdots \rangle'$
indicates that the momentum conserving delta function and a factor of $(2 \pi)^3$ are removed.
Here, the power spectrum is given by
\begin{equation}
P_h^{\lambda}(k)=\frac{2H_{\inf}^2}{\Mpl^2 k^3}
\mcT_{k}^2(\tau_0)
\tilde{\gamma}_{k}^2\, |k\tau_r|^{(3-2\nu)}\frac{\tau_m}{\tau_r},
\end{equation}
where the two polarizations are not summed over and $P_h(k)=2P_h^\lambda(k)$
in our model.

From the definition of tensor bispectrum shown in Eq.~\eqref{uncBdef},
one finds
\begin{align}
\langle  h^{\lambda_1}_{\bm{k}_1}h^{\lambda_1}_{\bm{k}_2} h_{\bm{k}_3}^{\lambda_3}  \rangle^{\prime}
&=
e_{ij}^{\lambda_1*}(\hat{\bm{k}}_1)e_{kl}^{\lambda_1*}(\hat{\bm{k}}_2)e_{mn}^{\lambda_3*}(\hat{\bm{k}}_3)
B_{ijklmn}(k_1,k_2,k_3)
\notag\\
&\simeq \frac{192}{a^3\Mpl^4}\mathcal{I}^{\rm (mass)}
\,
e_{ij}^{\lambda_1*}(\hat{\bm{k}}_1)e_{kl}^{\lambda_1*}(\hat{\bm{k}}_2)e_{mn}^{\lambda_3*}(\hat{\bm{k}}_3)\mathcal{E}^{\rm (mass)}_{ijklmn}(\hat{\bm k}_1,\hat{\bm k}_2,\hat{\bm k}_3)
\notag\\
&=\frac{192}{a^3\Mpl^4}\mathcal{I}^{\rm (mass)}
\,
e_{ij}^{\lambda_1*}(\hat{\bm{k}}_1)e_{jk}^{\lambda_1*}(\hat{\bm{k}}_2)e_{ki}^{\lambda_3*}(\hat{\bm{k}}_3),
\end{align}
where we ignored the subdominant contribution from the GR interaction in the second line.
This equation can be  reduced by summing over $\lambda_1$ in the squeezed limit, 
\begin{align}
\lim_{k_3 \to 0}\sum_{\lambda_1}\langle  h^{\lambda_1}_{\bm{k}_1}h^{\lambda_1}_{\bm{k}_2} h_{\bm{k}_3}^{\lambda_3}  \rangle^{\prime}
&\simeq \lim_{k_3 \to 0} \frac{-192}{a^3\Mpl^4}\mathcal{I}^{\rm (mass)}e_{ij}^{\lambda_3*}(\hat{\bm{k}}_3)
\hat{k}_1^i \hat{k}_1^j
\notag\\
&=\lim_{k_3 \to 0}\frac{-3\pi^{3/2}g_{\inf}}{2H_{\inf}\Mpl k_1^{3/2}}P_h(k_1)\sqrt{P_h^{\lambda_3}(k_3)}
\mathcal{J}^{\rm (mass)}e_{ij}^{\lambda_3*}(\hat{\bm{k}}_3)
\hat{k}_1^i \hat{k}_1^j,
\end{align}
where $\sum_\lambda e^{\lambda*}_{ij}(\hat{\bm k})e^{\lambda}_{jl}(\hat{\bm k})=\delta_{il}-\hat{k}_i \hat{k}_l$ is used in the first line.

Substituting it into Eq.~(\ref{expression_powerspectrum_tensor_ipoltensor}),
the observable power spectrum taking into account the presence of the long  tensor mode $h^{\lambda_3} _{\bm{k}_3}$
has the following quadrupole modulation 
\bea
\lim_{k_3 \to 0} P_h (\bm{k}_1) \bigr|_{h^{\lambda_3} _{\bm{k}_3}} ^{({\rm obs})}  &\simeq&
\lim_{k_3 \to 0}  P_h (k_1)
\left(1+ \hat{f}_{\rm NL}^{\lambda_3} (k_3) h^{\lambda_3} _{\bm{k}_3} e^{\lambda_3*} _{ij} (\hat{\bm{k}}_3)
\hat{k}_1 ^i \hat{k}^j _1\right) \nonumber\\
& \equiv&\lim_{k_3 \to 0}P_h (k_1) \left(1 + \tilde{\mathcal{Q}}^{\lambda_3} _{ij} (\bm{k}_3)
\hat{k}_1 ^i \hat{k}^j _1 \right)\,,
\ena
with
\begin{equation}
\hat{f}_{\rm NL}^{\lambda_3}(k_3)\equiv
\frac{-3\pi^{3/2}g_{\inf}}{2H_{\inf}\Mpl k_1^{3/2}}
\frac{\mathcal{J}^{\rm (mass)}}{\sqrt{P_h^{\lambda_3}(k_3)}}\,.
\end{equation}

Then, the observed quadrupole in the tensor power spectrum can be obtained by 
summing over both long-mode polarizations $\lambda_3=\pm$ and wave vector $\bm k_3$,
\begin{align}
\mathcal{Q}_{ij} ( \bm{x}) =  \lim_{k_3 \to 0}\int \frac{d^3 k_3}{(2 \pi)^3}  e^{i \bm{k}_3 \cdot \bm{x}} 
\sum_{\lambda_3} \tilde{\mathcal{Q}}^{\lambda_3} _{ij} (\bm{k}_3)\,.
\end{align}
Note that the lower limit of the integral in the above equation is given by the scale corresponding to the infrared cutoff which is, however,  not relevant in our case.
The expectation value of the quadrupole moments squared can be calculated as
\begin{eqnarray}
\left( \mathcal{Q} \right)^2 &\equiv& \frac{8 \pi}{15} 
\langle \mathcal{Q}_{ij}  (\bm{x}) \mathcal{Q}_{ij} (\bm{x})  \rangle
=\lim_{k_3 \to 0} \frac{8 \pi}{15} \int \frac{d^3 k_3}{(2 \pi)^3} \sum_{\lambda_3} 
\langle  \tilde{\mathcal{Q}}^{\lambda_3 *} _{ij} (\bm{k}_3)  \tilde{\mathcal{Q}}^{\lambda_3} _{ij} (\bm{k}_3) \rangle^{\prime}
\nonumber\\
&=&\lim_{k_3 \to 0}  \frac{4}{15 \pi} \int 
\dd k_3\, k_3^2  \sum_{\lambda_3}
\left(\hat{f}_{\rm NL} ^{\lambda_3} (k_3) \right)^2 
P_h^{\lambda_3}(k_3)
\nonumber\\
&=&\frac{6 \pi^2}{5}\left(\frac{g_{\inf}}{H_{\inf}\Mpl}\right)^2
\int\dd\kappa_3\, \kappa_3^2
\left(\mathcal{J}^{\rm (mass)}\right)^2. 
\end{eqnarray}

%
\begin{figure}[tbp]
  \begin{center}
  \includegraphics[width=100mm]{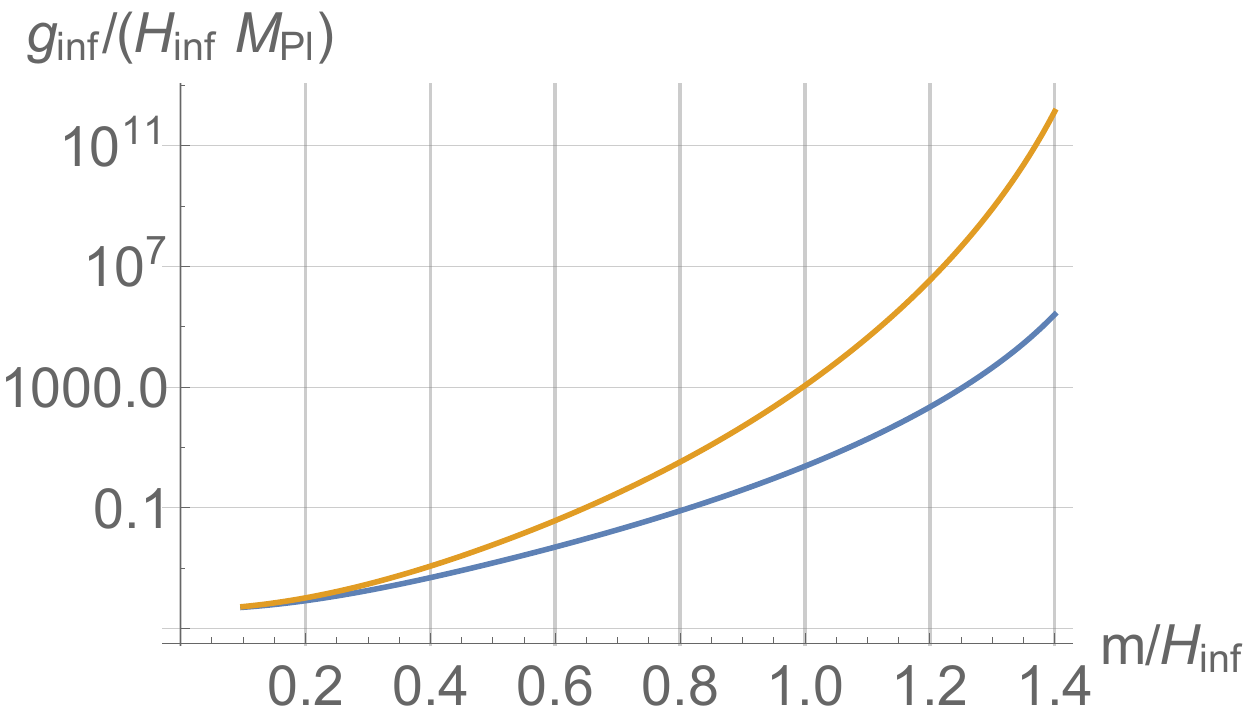}
  \end{center}
  \caption{Plots of $g_{\rm inf}/ (H_{\rm inf} M_{\rm Pl})$ that gives $\sqrt{( \mathcal{Q} )^2} = 10^{-2}$
  as functions of $m/H_{\rm inf}$.
  The color scheme is orange for $\kappa_{\rm UV}=10^{-15}$ and blue for $\kappa_{\rm UV}=10^{-8}$, which roughly correspond to LISA and SKA, respectively.
The lower bound of the $\kappa_3$ integral is taken as $10^{-15}\kappa_{\rm UV}$.}
 \label{ginf}
\end{figure}
%
Using Eq.~\eqref{squeezed J}, this integral is evaluated as
\begin{equation}
\int^{\kappa_{\rm UV}}\dd \kappa_3 \kappa_3^2
\left[\mathcal{J}^{\rm (mass)}\right]^2
\simeq \mathcal{G}\left(\frac{m}{H_{\inf}}\right)\, \frac{\kappa_{\rm UV}^{3-2\nu}}{3-2\nu},
\label{K def}
\end{equation}
with
\begin{align}
\mathcal{G}\left(\frac{m}{H_{\inf}}\right)&\equiv 
\left( 
\frac{2^\nu\Gamma(\nu)}{\pi}\,
\int^{\infty}_{|k_1 \tau_r|} \dd y \,
y^{\frac{3}{2}-\nu\pm 1}\, {\rm Im}
\left[ \left(H_\nu^{(1)}(y)\right)^2\right]\right)^2\notag\\
&\approx\exp\left[
1.80919\left(\frac{m}{H_{\inf}}\right)^3-2.01653\left(\frac{m}{H_{\inf}}\right)^2-4.96751\left(\frac{m}{H_{\inf}}\right)+5.58025\right],
\label{mathcalg_exp}
\end{align}
where $\kappa_{\rm UV}$ denotes UV cutoff of $\kappa_3$ and a numerical fit is used in the second line.
It is worth mentioning that the concrete value of $\kappa_{\rm UV}$  depends on the target scale at which 
we expect to observe the quadrupole modulation. For example, by assuming that $k_3$ is slightly larger than
the horizon scale today, for the scale corresponding to interferometers like LISA, $\kappa_{\rm UV} \simeq 10^{-15}$,
while for the one corresponding to pulsar timing arrays like SKA,   $\kappa_{\rm UV} \simeq 10^{-8}$.

Combining all results in this section so far, $g_{\rm inf}$ can be expressed as%
 \footnote{For the present case with $m/H_{\rm inf} = \mathcal{O} (1)$, a theoretically natural choice for the value of $g_{\rm inf}$ would be of order $\mathcal{O}(H_{\rm inf}^2)$, which can be different from $\mathcal{O}(H_{\rm inf} \Mp)$. Moreover, theoretical considerations such as the positivity bound may lead to some constraints on the range of $g_{\rm inf}$. However, in the present paper, we instead adopt the purely phenomenological standpoint and discuss the interesting range of $g_{\rm inf}$ for on-going/future experiments. For this purpose it is convenient to put bounds on $g_{\rm inf}$ in the unit of $H_{\rm inf} \Mp$ instead of $H_{\rm inf}^2$, assuming that $H_{\rm inf}$ is not too different from $\Mp$.}
\begin{align}
\frac{g_{\rm inf}}{H_{\rm inf} \Mpl} = \sqrt{\frac{5}{6 \pi^2}} \sqrt{( \mathcal{Q} )^2}
\left(\frac{3-2 \nu}{\kappa_{\rm UV}^{3-2 \nu}}\right)^\frac12  
\mathcal{G}\left(\frac{m}{H_{\inf}}\right)^{-\frac12},
\label{g_inf_est}
\end{align}
According to recent discussions on the quadrupole modulation
induced by the tensor bispectrum, it was suggested that
the quadrupole can be observed if $\sqrt{( \mathcal{Q} )^2} \gtrsim 10^{-2}$ 
\cite{Dimastrogiovanni:2018uqy, Ozsoy:2019slf} on small scales,
where the tensor power spectrum is observable.
In Fig.~\ref{ginf}, numerical plots of $g_{\rm inf} / (H_{\rm inf} M_{\rm Pl})$
that give $\sqrt{( \mathcal{Q} )^2} = 10^{-2}$  are shown. 
If the value of $g_{\rm inf}$ for given $m/H_{\rm inf}$ 
is greater than the one shown by the plots, the modulation can be proved at the interferometer scale 
(the orange plot with $\kappa_{\rm UV} = 10^{-15}$) or the scale of pulsar timing arrays (the blue plot with $\kappa_{\rm UV} = 10^{-8}$).

Making use of the expression (\ref{g_inf_est}), the dependence of $g_{\rm inf} / (H_{\rm inf} M_{\rm Pl})$
on $\kappa_{\rm UV}$ and $m/H_{\rm inf}$ can be understood as follows.
For a fixed $m/H_{\rm inf}$, its power on $\kappa_{\rm UV}$ is $-3/2+\nu$ that is negative for $0< \nu < 3/2$.
Then, for $\kappa_{\rm UV}$ that is positive and much smaller than $1$, a larger 
$g_{\rm inf} / (H_{\rm inf} M_{\rm Pl})$ is required for the detectability with a smaller $\kappa_{\rm UV}$.
On the other hand, for a fixed $\kappa_{\rm UV}$, from the second line of (\ref{mathcalg_exp}),
we can show that $(\mathcal{G} (m/H_{\rm inf}))^{-1/2}$ is an increasing function of $m/H_{\rm inf}$ for the region
shown in Fig.~\ref{ginf}, which means that a larger $g_{\rm inf} / (H_{\rm inf} M_{\rm Pl})$ is required for the detectability with a larger $m/H_{\rm inf}$.
Qualitatively, both of these features can be explained by the fact that the tensor power spectrum is blue-tilted, 
where the effect of the super horizon mode crutial for the modulation is suppressed for smaller $\kappa_{\rm UV}$ 
with fixed $m/H_{\rm inf}$ or for larger  $m/H_{\rm inf}$ with fixed $\kappa_{\rm UV}$.
Then, for large $m/H_{\rm inf}$, the detectability requires a very large value of $g_{\rm inf} / (H_{\rm inf} M_{\rm Pl})$, which needs an enormous fine-tuning, but for small $m/H_{\rm inf}$, we do not need such a fine-tuning.
For example, for $m/H_{\rm inf}=0.5$, which still gives interesting results on the tensor power specturm \cite{FKMM},
in order to obtain  $\sqrt{( \mathcal{Q} )^2} \gtrsim 10^{-2}$, observable at the interferometer scale,
we just need
\begin{equation}
\frac{g_{\inf}}{H_{\inf}\Mpl}
\gtrsim 10^{-2}.
\end{equation}
%

\section{Conclusions and Discussions}
\label{sec:conclusions}

Recently, based on the minimal theory of massive gravity (MTMG), we proposed a new scenario
predicting blue-tilted and largely amplified primordial gravitational waves (GWs)~\cite{FKMM}.
In the scenario, the primordial GWs can be detected by interferometer experiments, even if their signal is not observed at the CMB 
scale. While the analysis in Ref.~\cite{FKMM} was limited to
the linear perturbation related with the tensor power spectrum, since there are many other possible 
sources producing the GWs detectable at the interferometer scales, it is important to clarify how to distinguish our scenario from others. In this paper, as a natural extension of the previous analysis, 
we have considered the non-Gaussianity of primordial GWs in the scenario with the special emphasis on
the tensor bispectrum.

We have shown that in MTMG, the interaction Hamiltonian for the tensor perturbation at the third order 
has two contributions, where one has the same form as the usual one derived from general relativity (GR)
and the other is peculiar to MTMG. With this interaction Hamiltonian,
we have calculated the tensor bispectrum based on the in-in formalism.
Our method to obtain the tensor bispectrum today can be separated into the following two steps.
The first step is evaluating the time integral in the in-in formalism during inflation. 
At this step, since the form of the interaction Hamiltonian peculiar to MTMG is the same as the one
appearing in supersolid inflation, the calculation of this part itself is not new,
while we have presented new results for the parameter region with the graviton mass comparable 
to the Hubble expansion rate during inflation. The second step is taking into account the enhancement
of the GWs after inflation  that is crucial in our scenario and is new. Combing these together, we have found that
the contribution from the three-point interaction peculiar to MTMG dominates the one 
derived from GR and that the resultant tensor bispectrum peaks at the squeezed limit whose slope is determined
by the graviton mass.

We have also considered the detectability of this tensor bispectrum. Since it had been shown in the literature that the bispectrum cannot be probed directly at interferometer scales, we have instead discussed the detectability based on the quadrupolar modulation
of the tensor power spectrum, which is induced by the  squeezed tensor bispectrum,
making use of the well-known idea of the tensor fossils.
We have shown that for $m/H_{\rm inf} = 0.5$, which is sufficient to generate a blue-tilted and amplified tensor power spectrum
detectable at the interferometer scales,
if $g_{\inf}$, the coefficient of the three-point interaction peculiar to MTMG with the dimension of mass squared,
is larger than $10^{-2}   H_{\inf} \Mpl$, which is natural for $H_{\rm inf}$ not so smaller than $\Mpl$,
the quadrupolar modulation sourced by the squeezed tensor bispectrum is observable.

The appearance of the quadrupolar modulation of the
tensor power spectrum is related with the Maldacena's consistency relation on the squeezed limit
of the tensor three-point function~\cite{Maldacena:2002vr}
with which it was shown  that tensor modes with wavelengths much longer than the present Hubble radius
are unobservable~\cite{Pajer:2013ana}.
Actually, for the setup of solid inflation, in Ref.~\cite{Bordin:2016ruc}, 
it was shown that the appearance of the quadrupolar modulatoin 
is related with the fact that the Maldacena's consistency relation is violated in the model \cite{Endlich:2013jia, Akhshik:2014bla}.%
\footnote{For recent discussions on the implication of the violation of Maldacena's consistency relation 
in solid inflation, see~\cite{Bordin:2017ozj, Pajer:2019jhb}.}
It is interesting to see if the consistency relation is violated or not in the scenario we considered
whose setup can be regarded as generalization of solid inflation, explicitly. 

In this paper, we have restricted ourselves to the case where 
the massive graviton is the only spin-2 particle. On the other hand, recently,
from the viewpoint of cosmological collider~\cite{Arkani-Hamed:2015bza,Lee:2016vti},
the possibility that there are extra new particles in the very high energy regime like during inflation
has been actively explored. Although most of works so far are intended to find particles with spin less than 2,
some phenomenology is investigated for the case with extra spin-2
particles~\cite{Dimastrogiovanni:2018uqy,Kehagias:2017cym,Biagetti:2017viz,Franciolini:2017ktv,Franciolini:2018eno,Bordin:2018pca,Goon:2018fyu,Anninos:2019nib,Bordin:2019tyb,Dimastrogiovanni:2019bfl}. The generalization of the current work
to this direction might be worth investigating.

Finally, in this paper, as an extension of \cite{FKMM}, where the amplitude of tensor power spectrum 
is large at the interferometer scales, but small at the CMB scale, we have not considered
the possibility that the tensor bispectrum in this model is detectable by on-going CMB experiments.
However, for some parameter region, it is possible that the amplitude of the tensor bispectrum
is sufficiently large at the CMB scale, while that of the tensor power spectrum is not.
So far, the detectability of tensor bispectra by CMB experiments are discussed for 
only very limited types of the tensor bispectra whose forms are well approximated by given templates
(see \cite{Shiraishi:2019yux}, for a review) and the tensor bispectrum generated in our model
does not fall into such classes. Therefore, it might be  also interesting
to consider the detectability of the tensor bispectrum
in this model by CMB experiments. We would like to leave these topics for future work.

\acknowledgments

We would like to thank 
E. Dimastrogiovanni,
P. Creminelli, M. Fasiello, S. Koroyanagi, 
V. De Luca, G. Franciolini, A. Ricciardone and G. Tasinato for useful discussions.
The work of TF was supported by JSPS KAKENHI
No. 17J09103 and No. 18K13537.
The work of SMi was supported by JSPS KAKENHI
No. 16K17709.
The work of SMu was supported by
JSPS KAKENHI  No. 17H02890,
No. 17H06359 and also
partially supported by the World Premier International Research Center Initiative (WPI Initiative), MEXT, Japan.

\appendix
\section{Graviton mass and coupling constant in MTMG Theory \label{appendix_MTMG}}

In the main text, we have studied primordial tensor non-Gaussianity from massive gravity
which predicts blue-tilted and largely amplified gravitational waves without relying on details of 
a concrete theory. On the other hand, MTMG \cite{DeFelice:2015hla, DeFelice:2015moy},
which has a mass scale $m$ and three dimensionless parameters $c_i$ $(i=1,2,3)$,
is a concrete example giving such interesting phenomenology.
Therefore,  here, we express
the tensor graviton mass $\mu$ and coupling constant $g$ in terms of the parameters in MTMG.
The FLRW cosmology in this theory has two branches of solutions, the self-accelerating branch
and the normal branch. In the self-accelerating branch, the effective cosmological constant is given by
\bea
\Lambda_{\rm eff} = \frac{m^2}{2} X (c_1 X^2 + 3 c_2 X + c_3)\,,
\ena
where  $X$ is a constant satisfying $c_1 X^2 + 2 c_2 X + c_3 =0$. 
In this set-up, it was shown that the squared mass of graviton is given by
\bea
\mu^2 = \frac{m^2}{2}X \left[c_2 X + c_3 + \frac{H}{H_f} (c_1 X + c_2) \right]\,,
\label{graviton_mass}
\ena
where $H_f$ is the Hubble expansion rate of the fiducial metric and
it can be freely specified. 
The self-coupling constant of the cubic interaction is given by
\bea
g = -\frac{m^2}{24} X \left[c_2 X - c_3 + \frac{H}{H_f} (c_1 X -c_2) \right]\,.
\label{g in MTMG}
\ena

\section{Real Part versus Imaginary Part}
\label{Real Part versus Imaginary Part}

In this section, we derive the approximation in Eqs.~\eqref{I GR} and \eqref{I mass}. The integrals in Eqs.~\eqref{IGR def} and \eqref{Imass def} are evaluated at the end of inflation as \begin{align}
\int^{\tau_r}_{-\infty} \dd \eta\,  a^{\mp1}(\eta)  {\rm Im}\left[v_{k_1}^*(\tau)v_{k_2}^*(\tau)v_{k_3}^*(\tau)
v_{k_1}(\eta)v_{k_2}(\eta)v_{k_3}(\eta)\right]=\mathcal{I}_{\rm im}+\mathcal{I}_{\rm re},
\end{align}
with
\begin{align}
\mathcal{I}_{\rm im}&={\rm Im}\left[v_{k_1}^*(\tau_r)v_{k_2}^*(\tau_r)v_{k_3}^*(\tau_r)
\right]
\int^{\tau_r}_{-\infty} \dd \eta\,  a^{\mp1}(\eta)  {\rm Re}\left[
v_{k_1}(\eta)v_{k_2}(\eta)v_{k_3}(\eta)\right],
\notag\\
\mathcal{I}_{\rm re}&={\rm Re}\left[v_{k_1}^*(\tau_r)v_{k_2}^*(\tau_r)v_{k_3}^*(\tau_r)
\right]
\int^{\tau_r}_{-\infty} \dd \eta\,  a^{\mp1}(\eta)  {\rm Im}\left[
v_{k_1}(\eta)v_{k_2}(\eta)v_{k_3}(\eta)\right].
\end{align}
Since its integrand increases in time, 
$\mathcal{I}_{\rm re}$ is analytically performed with the super-horizon asymptotic form of the mode function as
\begin{align}
\mathcal{I}_{\rm re}
&={\rm Re}\left[v_{k_1}^*(\tau_r)v_{k_2}^*(\tau_r)v_{k_3}^*(\tau_r)
\right]
\int^{\tau_r}_{-\infty} \dd \eta\,  a^{\mp1}(\eta)  {\rm Im}\left[
v_{k_1}(\eta)v_{k_2}(\eta)v_{k_3}(\eta)\right]
\notag\\
&={\rm Re}\left[v_{k_1}^*(\tau_r)v_{k_2}^*(\tau_r)v_{k_3}^*(\tau_r)
\right]
\frac{2^{3(\nu-1)}\Gamma^3(\nu)}{\pi^{3/2}k_1^{5/2}(\kappa_2\kappa_3)^\nu}
\left(\frac{k_1}{H_{\inf}}\right)^{\mp1}
\int_{|k_1\tau_r|}^{\infty} \dd y\,  y^{\frac{3}{2}\pm1-3\nu}
\notag\\
&\simeq{\rm Im}\left[v_{k_1}^*(\tau_r)v_{k_2}^*(\tau_r)v_{k_3}^*(\tau_r)
\right]
\frac{2^{2\nu-3}\pi^{3/2}\Gamma^2(\nu)}{k_1^{5/2}(\kappa_2\kappa_3)^\nu\Gamma(\nu+1)}
\frac{|k_1\tau_r|^{\frac{5}{2}-\nu}}{\frac{5}{2}\mp1-3\nu}a_r^{\mp1}
(1+\kappa_2^{2\nu}+\kappa_3^{2\nu}),
\end{align}
where $m/H_{\inf}<0.94$ is assumed to simplify $\int \dd y y^{-3\nu+5/2}$.
Using Eq.~\eqref{Re integ}, one finds
\begin{equation}
\frac{\mathcal{I}_{\rm re}}{\mathcal{I}_{\rm im}}\simeq
\frac{2^{2\nu}\Gamma^2(\nu)(1+\kappa_2^{2\nu}+\kappa_3^{2\nu})}{\Gamma(\nu+1)(\kappa_2\kappa_3)^\nu \,\mathcal{J}^{\rm (GR/mass)}}\,
\frac{|k_1\tau_r|^{\frac{5}{2}-\nu\pm1}}{\frac{5}{2}\mp1-3\nu}.
\end{equation}
The factor $|k_1\tau_r|^{\frac{5}{2}-\nu\pm1}$ is tiny in the massive case.
Therefore, this shows $\mathcal{I}_{\rm re}\ll \mathcal{I}_{\rm im}$
which justifies the approximation in Eqs.~\eqref{I GR} and \eqref{I mass}.


\end{document}